\documentclass[10pt,twocolumn,twoside]{IEEEtran}
\IEEEoverridecommandlockouts

\usepackage{cite}
\usepackage{amsmath,amssymb,amsfonts}
\usepackage{algorithmic}
\usepackage{graphicx}
\usepackage{textcomp}
\usepackage{balance}
\usepackage{xcolor}
\usepackage[utf8]{inputenc}

\usepackage{comment}

\usepackage{multicol}
\usepackage{bm}
\usepackage{amsmath,amssymb,amsfonts}
\usepackage{amsthm}
\usepackage{tabularx}
\usepackage{makecell}
\usepackage{multirow}
\usepackage{booktabs}
\usepackage[hidelinks]{hyperref}
\usepackage{tikz}
\usepackage{pgfplots}
\usepgfplotslibrary{groupplots}
\pgfplotsset{compat=1.18}
\usetikzlibrary{positioning}
\usepackage{svg}

\usepackage{scalerel,stackengine,amsmath}
\newcommand\equalhat{\mathrel{\stackon[1.5pt]{=}{\stretchto{%
    \scalerel*[\widthof{=}]{\wedge}{\rule{1ex}{3ex}}}{0.5ex}}}}

\newtheorem{definition}{Definition}

\newtheorem{example}{Example}

\newtheorem{axiom}{Axiom}

\def\BibTeX{{\rm B\kern-.05em{\sc i\kern-.025em b}\kern-.08em
    T\kern-.1667em\lower.7ex\hbox{E}\kern-.125emX}}
\usepackage{lipsum}

\title{\LARGE \bf 
A Welfarist Perspective on Fair Generation Curtailment
}

\author{Jonas G. Matt, Ilia Shilov, and Saverio Bolognani
\thanks{The authors are with the Automatic Control Laboratory (IfA) at ETH Zurich, 8092 Zurich, Switzerland. Email: \{jmatt, ishilov, bsaverio\}@ethz.ch}%
\thanks{This work was supported by the Swiss Federal Office of Energy (grant SI/502734 MAESTRO) and by the Swiss National Science Foundation under the NCCR Automation (grant agreement 51NF40\textunderscore225155).}%
}

\begin{document}
\begingroup
\allowdisplaybreaks

\maketitle

\begin{abstract}

This paper presents a welfarist approach to fair active power curtailment in distribution grids with distributed photovoltaics. We address the lack of consistent axiomatic foundations in existing ad-hoc curtailment rules by modeling the decision as a social choice problem over feasible operating points and by deriving curtailment objectives from a set of foundational axioms that express principled stances on fairness and grid access rights. Rather than relying on the typically assumed full comparability of utilities, which can lead to undesirable outcomes in heterogeneous residential systems, we adopt a cardinal non-comparability stance on utilities. This approach requires far fewer assumptions about prosumers' private preferences while providing a rigorous basis for fair social ranking. We then present a unified framework that demonstrates that existing curtailment schemes represent specific instances of the Kalai-Smorodinsky rule applied to different normative reference points. This perspective offers grid operators an auditable, axiomatic foundation for justifying fairness in local energy systems.

\end{abstract}



\section{Introduction}

The proliferation of distributed photovoltaics (PV) in local energy systems necessitates active power curtailment to respect grid operational constraints (up to 10\% of the power generated by new installations \cite{NOVAN2024102930} and up to 8\% of the total solar generation \cite{OShaughnessy2020}).
Grid operators face the problem of deciding which generators to curtail, and this decision is guided by multiple (possibly conflicting) criteria:
\begin{itemize}
    \item \textbf{Sustainability}: the transition towards a sustainable power system calls for minimizing the total curtailment of renewable sources (in some cases, by law \cite{Schermeyer2018,EEG2017});
    \item \textbf{Fair grid access}: Distribution grid operators are often subject to regulations that mandate non-discriminatory access to the network \cite{EU2019_944};
    \item \textbf{Social equity}: society expects fair sharing of a public resource like the power grid \cite{Brockway2021,dewinkelAdaptingLimitedGrid2025a};
    \item \textbf{Investment promotion}: different curtailment schemes can affect profitability and therefore promote or discourage investments in renewable generation \cite{Cuenca2023}.
\end{itemize}
All these criteria contribute to the definition of what can be called \emph{fair generation curtailment}.

The current literature addresses this challenge by proposing a variety of rules \cite{gebbran2021fair, lusis2019reducing, liu2020fairness,Borbath2024,Alam2024,Petrou2021,vadavathiFairEfficientCongestion2024}. Few of them have an axiomatic justification, and their ad-hoc nature can lead to undesirable outcomes (such as loss of efficiency and disincentivization of future investments) and little transparency \cite{stringer2021fair}.
This variety reflects distinct fairness intuitions, yet it leaves designers without a clear procedure to choose a metric and to justify the corresponding fair objective \cite{soares2024review,dewinkelReviewFairnessConceptualizations2024}.
Without a principled approach, using such metrics can lead to undesirable outcomes. 
For example, metrics like Jain fairness or the Gini index can favor curtailment decisions that are not Pareto optimal (that is, there exists another decision that would have been preferred by all involved parties), leading to system-wide inefficiency \cite{chen2023guide}.
The core issue is that these methods lack a consistent axiomatic foundation for what constitutes a ``fair'' preference between curtailment decisions.
This paper argues that a generalized and principled approach is required, which provides a rigorous and explainable way to choose among various specific fairness metrics. 

We take a \emph{welfarist} stance \cite{shilov2025welfare}, which views the curtailment decision as a social choice over feasible operating points, informed by prosumers' welfare.
We rely on key results from social choice theory \cite{Roberts1980} that state that the optimal decision must be determined via a social welfare function, whose form follows from a set of basic axioms and from a deliberate decision on how commensurable the utilities of different agents (interpersonal comparability) are.
In this paper we make this connection operational for energy curtailment, in a way that is straightforward to implement in an optimal power flow (OPF) setting.
We show that the only decision rule that meets the specifications of this application is the minimization of the Kalai-Smorodinsky (KS) social welfare function \cite{Kalai1975}.

\begin{figure}[t!]
    \centering
    \includegraphics[width=\linewidth]{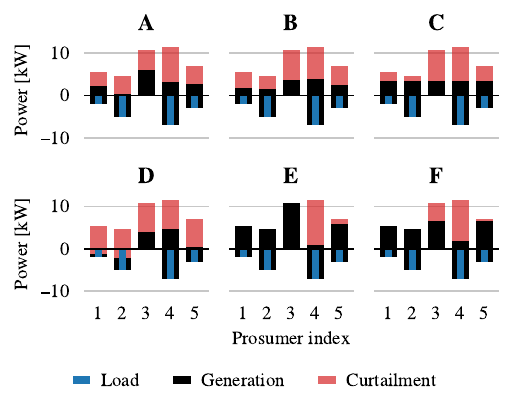}
    \caption{All these curtailment profiles result in marginally feasible voltages. The portion of generation above the zero line is exported power. Which one should a grid operator choose? The schemes underlying the profiles A-F are described in later sections of this paper.}
    \label{fig:comparison-curtailments}
\end{figure}

By deriving social welfare functions from foundational axioms, we enable designers to tune policy parameters to align with specific regulatory or social objectives. 
The design parameters that need to be chosen are particularly easy to interpret and defend: namely, they amount to the definition of what levels of grid access the prosumers are entitled to, and the grid access that the prosumers would enjoy if they were not sharing the grid with others. 
This shifts the focus from applying arbitrary fairness indices (with unintended inefficiency and inequality consequences) to making explicit, auditable design choices grounded in a rigorous social choice framework. Interestingly, many curtailment criteria proposed in the literature (for example, some of those reviewed in \cite{gebbran2021fair,liu2020fairness}) can be explained and justified via this axiomatic approach. 

The remainder of this paper is organized as follows: Section \ref{sec:curtailment} formalizes the curtailment decision problem and defines the feasible operating set. Section \ref{sec: welfarism} establishes the welfarist framework, detailing the foundational axioms and the derivation of the KS social welfare function. Section \ref{sec:unification} unifies existing curtailment schemes from the literature by demonstrating that they represent specific configurations of the KS rule. Section \ref{sec: simulations} evaluates these approaches through time-series simulations on a 6-bus low-voltage power grid testbed. Finally, Section \ref{sec: conclusion} concludes the paper.
\section{The curtailment decision problem}
\label{sec:curtailment}

We consider a static curtailment decision at a specific time instant on a distribution feeder with $N$ prosumers. Each prosumer $i$ has a local PV generation $p_i$ and a power demand $d_i$.
Because of voltage, line current, and transformer limits, the grid may not support the concurrent injection of the available active power from all prosumers.
The operator must therefore set an \emph{operating envelope} $x_i$ for each prosumer, understood as the maximal active power that the prosumer is allowed to generate at that time.
The following relation holds:
\[
\forall i: \quad 0 \le p_i \le x_i \le \bar p_i,
\]
where
$\bar p_i$ denotes the potential power generation at that specific time.
We denote by $p$, $x$, $\bar p$, the vectors obtained by stacking the respective quantities for all $N$ prosumers.
We represent the grid constraints via the nonlinear scalar inequality
\[
h(p) \le 0.
\]
The set of all admissible envelopes is therefore
\[
\mathcal{X}\ :=\ \big\{\,x\in\mathbb{R}^N:\ h(x)\le 0,\ 0\le x\le \bar p\,\big\}.
\]
In this work, we make a classical assumption: if a power generation vector $x$ is feasible, then any $y$ such that $0 \le y \le x$ is also feasible, allowing the system operator to freely curtail any prosumer down to the reference point without violating grid constraints (that is, $h(x)$ is convex and monotonic).\par
A \emph{curtailment scheme} is a selection rule that picks $x\in\mathcal{X}$.
It can be represented as an OPF problem
\begin{equation}\label{eq:curtailment}
    \begin{aligned}
        \text{optimize} \quad & W(x)\\
        \text{subject to} \quad & x \in \mathcal{X},
    \end{aligned}
\end{equation}
for some welfare function $W: \mathcal{X} \to \mathbb{R}$.
The curtailment profiles from Figure \ref{fig:comparison-curtailments} all correspond to different choices of $W$.

\section{Welfarist curtailment}\label{sec: welfarism}

We assume that agent (prosumer) $i$ derives a certain utility from power generation, where larger values are preferred.
For example, this utility may represent both the benefit of serving the local demand with local generation and the remuneration for exporting power to the grid.
In practice, the true agent utilities are not directly measurable. 
Therefore, a proxy for the utility of a given operating envelope $x\in\mathcal X$ for each agent $i$ must be selected, for example, generated power or power export.
We collect such a measure in a \emph{utility proxy metric} $u_i:\mathcal X\to\mathbb R$, we interpret $u_i(x)$ as the utility of agent $i$ at envelope $x$, and we define the \emph{utility profile}
\[
u(x) \;=\; \big(u_1(x),\dots,u_N(x)\big)\in\mathbb R^N.
\]
The set
\[
\mathcal{U} \;=\; \{\,u(x)\in\mathbb R^N: x\in\mathcal X\,\}
\]
represents the utility profiles that can be attained with a feasible decision (an admissible envelope). Selecting an operating envelope $x \in \mathcal{X}$ in \eqref{eq:curtailment} is a social choice problem: the designer must turn the profile $u(x)$ into a social ranking over feasible envelopes and, ultimately, into an objective for \eqref{eq:curtailment}. A social welfare function (SWF) $W(x)$ then represents the social ranking. Classic results \cite{Roberts1980, dAspremontGevers2002, Sen1979} show that the admissible form of such a function is largely determined by the basic welfarist axioms and the level of interpersonal comparability (that is, how we compare the utilities of different prosumers) the designer is willing to assume.


\subsection{Welfarism and basic axioms}

In what follows, we adopt a welfarist viewpoint: once utilities and the feasible set are fixed, all information that matters for the decision is contained in the utility vector at each envelope. 
On the basis of that information, we need to construct a social preference relation over the possible curtailment decisions.

\begin{definition}
    A \emph{social preference relation} $\succeq_u$ is a complete and transitive ranking of envelopes in $\mathcal{X}$, given a utility profile $u(\cdot)$.
\end{definition}
For $x,y\in \mathcal{X}$, the expression $x \succeq_u y$ is interpreted as ``society weakly prefers $x$ to $y$ under profile $u$''.
For the social preference to be consistent, we first specify which basic properties (or axioms) it needs to satisfy: Weak Pareto (WP), Independence of Irrelevant Alternatives (IIA), and Continuity (C), stated here informally. We refer to \cite{shilov2025welfare} for a formal treatment and discussion.
\begin{description}
    \item[WP]  If $u_i(x)>u_i(y)$ for all $i$, then $x\succ_u y$. WP guarantees ``efficiency'' of the allocation and rules out some SWF that might not fully distribute available utility, see Remark \ref{remark: WP}.
    \item[IIA] The ranking between $x$ and $y$ depends only on $\big(u_1(x),\dots,u_n(x)\big)$ and $\big(u_1(y),\dots,u_n(y)\big)$. 
    \item[C] For any $x \succ y$, there exists $\varepsilon > 0$ such that $|u(x') - u(x)| < \varepsilon$ and $|u(y') - u(y)| < \varepsilon$ imply $x' \succ y'$. Small changes in the utility vectors at $x$ and $y$ do not reverse strict ranking.
\end{description}

Despite their simplicity, these axioms are critical for social choice problems, as illustrated in the following example.

\begin{example}[Gini index and Pareto efficiency; Figure \ref{figremark: WP}] \label{remark: WP}
Consider two agents and two feasible envelopes with utilities
\(u(x) = (2,1)\) and \(u(y) = (1,1)\).
Envelope \(x\) Pareto dominates \(y\) because both agents are weakly better off and agent one is strictly better off.
The Gini index for two agents is
\(G(u) = |u_1-u_2|/(u_1+u_2)\),
so \(G(u(y)) = 0\) and \(G(u(x)) = 1/3\).
If one were to choose envelopes by minimizing the Gini index, \(y\) would be preferred to \(x\) even though \(x\) is a Pareto improvement for both agents. Similar conflicts with WP can arise with other common metrics, for example, variance or the Jain fairness index. Without adherence to WP, these metrics can result in the unnecessary curtailment of all agents.
\end{example}

An optional property one might want to impose is Anonymity (A), which requires equal treatment of agents who only differ by labels. It can be seen as a desirable fairness property, however some curtailment regulations prescribe pre-determined priorities across different agents (for example, \cite{Dolan2014,Haque2016, syrtseva}), so in practice Anonymity may not always be appropriate. 

\begin{description}
    \item[A] For any permutation \(\pi: \mathbb{N} \to \mathbb{N}\) of agents, the ranking given by $W$ remains the same.
\end{description}



Under (WP), (IIA) and (C), one can represent the social preference by a continuous social welfare function \cite{shilov2025welfare, Roberts1980}.

\begin{definition}[Social welfare function (SWF)]
A continuous function $W:\mathbb{R}^n\to\mathbb{R}$ is a social welfare function for the profile $u$ if, for all $x,y\in \mathcal{X}$,
\[
x \succeq_u y
\quad\Longleftrightarrow\quad
W\big(u(x)\big) \;\ge\; W\big(u(y)\big).
\]
\end{definition}

In other words, an SWF aggregates the single-agent utilities to represent a social ranking.
Welfarism therefore restricts the choice of the objective in the OPF \eqref{eq:curtailment} to functions of the form $W\big(u(x)\big)$.

\begin{figure}
\centering
\begin{tikzpicture}
\begin{groupplot}[
    group style={
        group size=2 by 1,
        horizontal sep=1cm,
        vertical sep=1cm
    },
    width=5cm,
    height=5cm,
    axis lines=left,
    grid=both,
    label style={font=\scriptsize},
    tick label style={font=\scriptsize},
    legend style={
        at={(1.1,-0.3)},
        anchor=north,
        legend columns=3,
        column sep=.2cm,
        font=\scriptsize,
        draw=black
    }
]

\definecolor{color1}{HTML}{1a5478} 
\definecolor{color2}{HTML}{d94a22} 
\definecolor{color3}{HTML}{6ca133} 
\definecolor{color4}{HTML}{62b2bf}
\definecolor{color5}{HTML}{d8a959}
\definecolor{grey}{HTML}{454545}

\nextgroupplot[
    xmin=0, xmax=1.1,
    ymin=0, ymax=1.1,
    xlabel={cumulative share of agents},
    ylabel={cumulative share of utility}
]


\addplot[color2, thick, forget plot] coordinates {(0,0) (0.5,0.5) (1,1)};
\addplot[only marks, mark=*, mark size=2.2, draw=color2, fill=color2, forget plot]
    coordinates {(0.5,0.5)};

\addplot[color1, thick, forget plot] coordinates {(0,0) (0.5,0.333333) (1,1)};
\addplot[only marks, mark=square*, mark size=2.2, draw=color1, fill=color1, forget plot]
    coordinates {(0.5,0.333333)};

\addlegendimage{grey, thick}
\addlegendentry{Pareto frontier}
\addlegendimage{only marks, mark=square*, mark size=2.2, draw=color1, fill=color1}
\addlegendentry{allocation A}
\addlegendimage{only marks, mark=*, mark size=2.2, draw=color2, fill=color2}
\addlegendentry{allocation B}

\nextgroupplot[
    xmin=0, xmax=3.3,
    ymin=0, ymax=3.3,
    xlabel={$u_1$},
    ylabel={$u_2$}
]

\addplot[grey, thick, forget plot] coordinates {(0,3) (3,0)};

\addplot[only marks, mark=*, mark size=2.2, draw=color2, fill=color2, forget plot]
    coordinates {(1,1)};
\addplot[only marks, mark=square*, mark size=2.2, draw=color1, fill=color1, forget plot]
    coordinates {(2,1)};

\addplot[->, thick] coordinates {(1,1) (1.92,1)};

\end{groupplot}
\end{tikzpicture}
\caption{Lorenz curves and Pareto efficiency for two allocations. \emph{Left}: allocation B with utilities $(1,1)$ lies on the equality line and has Gini index $0$, allocation A with utilities $(2,1)$ has Gini index $1/3$. \emph{Right}: feasible utility set and Pareto frontier, allocation A Pareto dominates allocation B.}
\label{figremark: WP}
\end{figure}
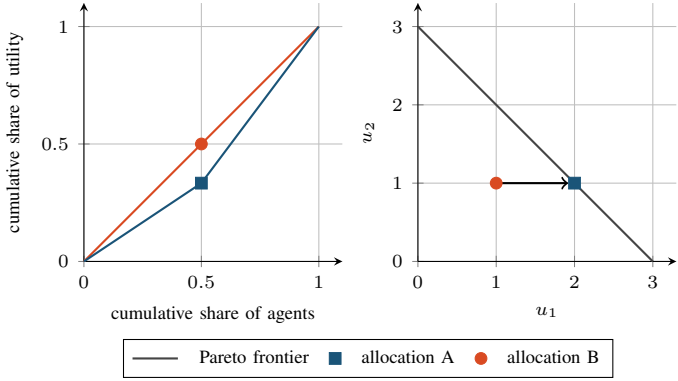

\subsection{Interpersonal comparability} 
\label{subsec:comparability}

The previous discussion says that the social ranking depends only on utility vectors through an SWF.
Classic results from social choice theory \cite{Roberts1980, dAspremontGevers2002, Sen1979} guide the choice of SWF.
They show that, for the aforementioned axioms to be satisfied, the admissible SWF is uniquely determined by the level of \emph{interpersonal comparability} one is willing to assume.
In other words, one must specify which statements involving utilities of different agents are meaningful.
We discuss two polar cases. 

\medskip

\subsubsection*{Cardinal full comparability (CFC)}

Under CFC, utilities are comparable both in levels and in increments across agents, which means that statements like
\[
u_i(x) - u_i(y) \;\ge\; u_j(z) - u_j(w)
\ \ \text{or}\ \ 
u_i(x) \;\ge\; u_j(y)
\]
are meaningful and that one unit of utility has the same significance for all agents. Under Weak Pareto, IIA, Continuity, and CFC invariance, admissible social welfare functions $W$ have representations of the form \cite{Roberts1980}

\begin{equation}
\label{eq:CFCform}
W\big(u(x)\big)
\;=\;
\frac{1}{n} \sum_{i\in N} u_i(x)
\;+\;
g\!\left(
\begin{bmatrix}
u_1(x)-\bar u(x)\\
\vdots\\
u_n(x)-\bar u(x)
\end{bmatrix}
\right),
\end{equation}
where $\bar u(x) = \tfrac{1}{n}\sum_{i\in N} u_i(x)$ is the average utility and $g:\mathbb{R}^n\to\mathbb{R}$ is homogeneous of degree one. The first term is utilitarian; the second term adjusts for inequality. Many familiar fairness metrics can be embedded in this class, for example a simple utilitarian-egalitarian mix with
\[
g(z) = \gamma \max_i z_i,
\qquad \gamma\in[0,1].
\]
Curtailment rule E from Figure \ref{fig:comparison-curtailments} is purely utilitarian, with $\gamma = 0$.

CFC is often implicitly assumed in wholesale energy markets, because one unit of energy is treated as comparable across agents, via its economic value.
For household-scale prosumers, this is hard to justify without detailed calibration of what utility energy brings to an individual customer.

\medskip

\subsubsection*{Cardinal non-comparability (CNC)}

Under CNC, utilities are cardinal for each agent but cannot be compared in levels or increments across agents. That can be formally expressed through the informational equivalence of utilities, that is, utilities $u_i$ and $u'_i$ express the same underlying preferences when they are transformed by \emph{agent-specific} positive affine maps:
\begin{equation}\label{eq:invariance}
    u'_i = a_i u_i + b_i,
\qquad a_i>0,\ b_i\in\mathbb{R}, \quad i\in N.
\end{equation}

Equivalence \eqref{eq:invariance} means that only statements that are invariant under individual rescalings and shifts are meaningful. One cannot say that agent $i$ values one unit of export more than agent $j$, or that a gain of two units for $i$ is better or worse than a gain of one unit for $j$. Only the direction of improvement for each agent and the ratios of utility gains are informative.

CNC expresses the informational stance we adopt for prosumer curtailment. We can compare envelopes for the same prosumer in cardinal terms, for example two extra kilowatt of export is better than one, but we do not assume that one kilowatt of export has the same utility for different households. Compared to CFC, CNC places far fewer assumptions on the utilities of agents.

A classical impossibility result due to Sen and others \cite{Sen1979} shows that, in general, there is no continuous social welfare function that satisfies Weak Pareto, IIA, Continuity, and CNC invariance for all utility profiles. In that sense, the CNC setting is too weak to admit a globally defined $W$ without additional structure.

However, when one allows a small amount of additional information in the form of reference envelopes (that is, relaxing IIA), a suitable welfare rule exists, as discussed in the next section.

\subsection{The Kalai-Smorodinsky social welfare function}
\label{subsec:NashKS}

We fix a utility profile $u$ and the feasible set $\mathcal{X}$. In addition to the basic assumptions (Weak Pareto, Continuity, and CNC invariance), we now relax IIA and allow the designer to use two reference envelopes.

\begin{itemize}
    \item \textbf{Fallback envelope:} We suppose that the designer can identify a feasible envelope $x^0 \in \mathcal{X}$ that serves as a common ``worst-case'' benchmark,
    for example $x_i = 0$ for all prosumers.
    We assume that comparisons are only made between envelopes that weakly Pareto improve on $x^0$.
    The fallback utility is $u^0 := u(x^0)$.
    \item \textbf{Utopia envelope:} We also suppose that the designer can identify a point $x^{\max} \in \mathcal{X}$ that corresponds to the coordinate-wise maximal utilities $u_i^{\max}(\mathcal{U}) \;:=\; u_i(x^{\max}) \;=\; \sup\{u_i(x) : x\in \mathcal{X}\}$, for all prosumers. 
    The utopia envelope may not be feasible for all agents simultaneously, but it represents the best that each agent could hope for.
\end{itemize}

With the two envelopes $x^0$ and $x^{\max}$ available, one can measure individual gains from the fallback and relative positions between fallback and utopia. This allows us to define the social ranking using the Kalai-Smorodinsky (KS) solution \cite{Kalai1975}.

\begin{definition}[KS solution]

A utility vector $u^{\mathrm{KS}} \in \mathcal{U}$ is a KS solution if:

\begin{enumerate}
    \item $u^{\mathrm{KS}}$ is Weakly Pareto efficient in $\mathcal{U}$
    \item there exists a common fraction $\lambda \in [0,1]$ such that 
    \begin{equation}
        \frac{u^{\mathrm{KS}}_i - u^0_i}{u^{\max}_i - u^0_i} = \lambda, \qquad \forall i \in N. \label{eq:KS_equalratio}
    \end{equation}
\end{enumerate}
\end{definition}

In an optimization context, this solution is identified by maximizing the following SWF:

\begin{equation}
\label{eq:KSW}
    W^{\mathrm{KS}}\big(u(x)\big) := \min_{i \in N} \frac{u_i(x) - u_i^0}{u_i^{\max} - u_i^0}.
\end{equation}

$W^{\mathrm{KS}}(u(x))$ represents the smallest relative gain from the fallback toward the utopia across all agents. 
Under the assumption of a comprehensive domain, the KS rule is the unique solution that satisfies Weak Pareto, CNC invariance, Continuity, Anonymity, and an additional axiom of Individual Monotonicity (IM), defined as follows:

\begin{description}
    \item[IM] If the feasible set expands from $\mathcal{U}_1$ to $\mathcal{U}_2$ such that agent $i$'s utopia point improves or remains the same, while all other agents' utopia points remain fixed ($u_j^{\max}(\mathcal{U}_2) = u_j^{\max}(\mathcal{U}_1)$ for $j \neq i$), then agent $i$ must not receive less utility: $u^{KS}_i(\mathcal{U}_2) \ge u^{KS}_i(\mathcal{U}_1)$, where $u^{KS}_i(\mathcal{U})$ explicitly denotes the KS solution for a specific feasibility set $\mathcal{U}$.
\end{description}



In our setting, \eqref{eq:KSW} corresponds to choosing envelopes that maximize the minimum relative improvement over the fallback, measured as a share of what would be attainable at utopia. The requirement for IM makes the KS rule particularly suited for promoting investments: if a prosumer improves their utopia point, for example investing in local generation upgrades, their utility should not decrease.

Another advantage of the KS solution is its computational simplicity: the requirement to find a common fraction $\lambda$ effectively turns the curtailment problem \eqref{eq:curtailment} into a single-parameter optimization. This makes the approach numerically robust and highly scalable for real-time grid operations.

Prominent CNC-invariant alternatives to the KS rule, such as the Nash product $W^{\mathrm{Nash}}(u) = \prod (u_i - u_i^0)$, satisfy the IIA$+1$ (that is, only one additional point is required) but violate IM (see Figure \ref{fig:NashKS}). In a curtailment context, a Nash-based rule (for example OF-Fair in \cite{Petrou2020}) could theoretically ``punish'' a prosumer by granting a smaller envelope following a grid expansion that was intended to benefit them, thus creating a disincentive for local generation upgrades.
Panel F in Figure \ref{fig:comparison-curtailments} corresponds to Nash-based curtailment of export.

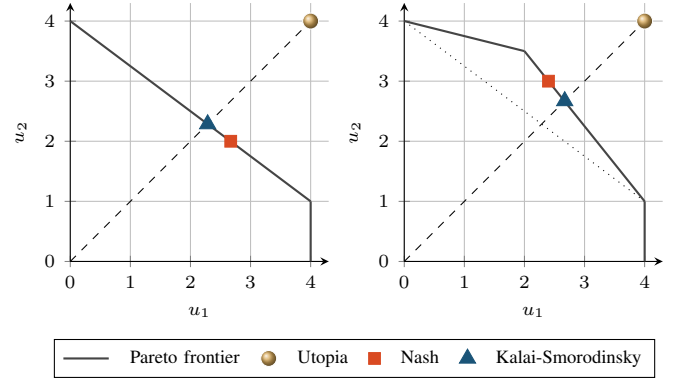
\begin{figure}[t]
    \centering
    \begin{tikzpicture}
\begin{groupplot}[
    group style={
        group size=2 by 1,   
        horizontal sep=1cm,
        vertical sep=1cm
    },
    width=5cm,
    height=5cm,
    axis lines=left,
    xmin=0, xmax=4.3,
    ymin=0, ymax=4.3,
    xlabel={$u_1$},
    ylabel={$u_2$},
    label style={font=\scriptsize},
    tick label style={font=\scriptsize},
    grid=both,
    legend style={
        at={(1.1,-0.3)},     
        anchor=north,
        legend columns=4,
        column sep=.2cm,
        font=\scriptsize
    }
]

\definecolor{color1}{HTML}{1a5478}
\definecolor{color2}{HTML}{d94a22}
\definecolor{color3}{HTML}{6ca133}
\definecolor{color4}{HTML}{62b2bf}
\definecolor{color5}{HTML}{d8a959}
\definecolor{grey}{HTML}{454545}


\nextgroupplot[
    legend entries={Pareto frontier,Utopia,Nash,Kalai-Smorodinsky},
    legend image post style={scale=1.0},
]

\addplot[grey, thick] coordinates {(0,4) (4,1)};
\addplot[only marks, mark=ball, mark size=3, ball color=color5, draw=white] coordinates {(4,4)};
\addplot[only marks, mark=square*, mark size=2.2, draw=color2, fill=color2] coordinates {(2.67,2)};
\addplot[only marks, mark=triangle*, mark size=3.5, draw=color1, fill=color1] coordinates {(2.286,2.286)};

\addplot[grey, thick] coordinates {(4,0) (4,1)};
\addplot[black, dashed] coordinates {(0,0) (4,4)};

\nextgroupplot[
]

\addplot[only marks, mark=ball, mark size=3, ball color=color5, draw=white] coordinates {(4,4)};
\addplot[only marks, mark=square*, mark size=2.2, draw=color2, fill=color2] coordinates {(2.4,3)};
\addplot[only marks, mark=triangle*, mark size=3.5, draw=color1, fill=color1] coordinates {(2.67,2.67)};

\addplot[black, dotted] coordinates {(0,4) (4,1)};
\addplot[grey, thick] coordinates {(0,4) (2,3.5)};
\addplot[grey, thick] coordinates {(2,3.5) (4,1)};
\addplot[grey, thick] coordinates {(4,0) (4,1)};
\addplot[black, dashed] coordinates {(0,0) (4,4)};

\end{groupplot}
\end{tikzpicture}
    \caption{Nash vs. KS. \textit{Left:} Original Pareto frontier and solutions for a two-agent problem. \textit{Right:} Expanded Pareto frontier. Nash moves to a point with lower \(u_1\) and higher \(u_2\), while KS moves up the fallback–utopia line and improves both utilities.}
    \label{fig:NashKS}
\end{figure}

\section{Existing fairness schemes as KS solutions}
\label{sec:unification}

We now revisit the main curtailment schemes proposed in the literature and show that each is a KS allocation for a specific choice of fallback and utopia points within the generation metric space $u_i(x) = x_i$. This unification demonstrates that these schemes all share the same underlying axioms (that could be viewed as the same fairness principles), but differ in their normative definitions of the ``fallback'' envelope (an envelope that we can guarantee to the prosumers regardless of the competing requests) and ``utopia'' envelope (what each prosumer would claim if there was no need to share the grid).
In all cases, the envelope $x$ is constrained by the feasible set $\mathcal{X}$.

\subsection{OPF Generation \texorpdfstring{\cite{liu2020fairness}}{liu2020}}\label{subsec: opf_generation}
Liu et al.\ \cite{liu2020fairness} equalize the fraction of available generation for each prosumer. They seek a scalar $\gamma^{G} \equalhat \lambda \in [0,1]$ such that $x_i = \gamma^{G} \bar{p}_i$. By setting the fallback to zero ($u^0_i = 0$) and utopia to available generation ($u^{\max}_i = \bar{p}_i$), the KS equality condition becomes:
\begin{equation*}
    \frac{u_i(x) - u^0_i}{u^{\max}_i - u^0_i}
    \;=\;
    \frac{x_i - 0}{\bar{p}_i - 0}
    \;=\;
    \frac{x_i}{\bar{p}_i}
    \;=\;
    \lambda,
    \qquad i \in N.
\end{equation*}
Maximizing $\lambda$ subject to feasibility yields the OPF Generation solution.
OPF Generation is panel A in Figure \ref{fig:comparison-curtailments}.
In \cite{zhanFairnessIncorporatedOnlineFeedback2024}, it is used in an online feedback optimization scheme.

\subsection{OPF Export \texorpdfstring{\cite{liu2020fairness}}{liu2020} and Proportional Redistribution \texorpdfstring{\cite{gebbran2021fair}}{gebbran2021}}\label{subsec: opf_export_proportional}
When local load $d_i$ is present, these schemes equalize export as a fraction of export capability ($\bar{p}_i - d_i$) \cite{liu2020fairness, gebbran2021fair}. By choosing the fallback as the local demand $u^0_i = d_i$ and utopia as available generation $u^{\max}_i = \bar{p}_i$, the KS ratio represents relative export gains:
\begin{equation*}
    \frac{u_i(x) - u^0_i}{u^{\max}_i - u^0_i}
    \;=\;
    \frac{x_i - d_i}{\bar{p}_i - d_i}
    \;=\;
    \lambda,
    \qquad i \in N.
\end{equation*}
This ensures that exporting prosumers share a common fractional progress from their load toward their individual potential. The schemes in \cite{liu2020fairness,gebbran2021fair} choose a parameter $\gamma^{E} \equalhat \lambda$ as large as feasibility permits, so the resulting curtailment is efficient in $\mathcal{U}$ and satisfies \eqref{eq:KS_equalratio} with maximal $\lambda$ for exporting prosumers.
Proportional redistribution is also selected in \cite{Liu2024} as the criterion for optimal selection of operating envelopes for distributed generators.
OPF Export is panel B in Figure \ref{fig:comparison-curtailments}.

\subsection{Uniform Dynamic Export \texorpdfstring{\cite{gebbran2021fair}}{gebbran2021}}\label{subsec: opf_export_uniform}
The uniform dynamic scheme introduces a common scalar cap on export at each time period \cite{gebbran2021fair}. We set the reference $u^0_i = d_i$ and introduce a normative utopia $u^{\max}_i = d_i + K$, where $K$ is a common export entitlement. For prosumers curtailed by the cap, the KS ratio is:
\begin{equation*}
    \frac{u_i(x) - u^0_i}{u^{\max}_i - u^0_i}
    \;=\;
    \frac{x_i - d_i}{(d_i + K) - d_i}
    \;=\;
    \frac{x_i - d_i}{K}
    \;=\;
    \lambda.
\end{equation*}
Since $K$ is constant across agents, equalizing $\lambda$ is equivalent to equalizing the export volume $x_i - d_i$ to a common value $\lambda K$.
Uniform Dynamic Export is panel C in Figure \ref{fig:comparison-curtailments}.

\subsection{Egalitarian Curtailment \texorpdfstring{\cite{gebbran2021fair}}{gebbran2021}}\label{subsec: opf_egalitarian}
Egalitarian schemes equalize the absolute curtailed volume $c_i = \bar{p}_i - x_i$ \cite{gebbran2021fair}. We frame this as a KS solution by setting the utopia to potential generation $u^{\max}_i = \bar{p}_i$ and the reference to $u^0_i = \bar{p}_i - \underline{c}$, where $\underline{c}$ is a common reference curtailment level. 
The KS ratio becomes:
\begin{equation*}
    \frac{u_i(x) - u^0_i}{u^{\max}_i - u^0_i}
    \;=\;
    \frac{x_i - (\bar{p}_i - \underline{c})}{\bar{p}_i - (\bar{p}_i - \underline{c})}
    \;=\;
    \frac{\underline{c} - c_i}{\underline{c}}
    \;=\;
    1 - \frac{c_i}{\underline{c}}
    \;=\;
    \lambda.
\end{equation*}
For a fixed $\underline{c}$, requiring a common $\lambda$ across agents directly entails that the curtailment $c_i$ is identical for all agents.
Notice that this perspective exposes an implicit assumption of the Egalitarian Curtailment approach: it must be possible to achieve a feasible state of the grid by applying equal curtailment to the prosumers (in other words, the fallback solution $x^0$ must be feasible), which is not guaranteed in general.
Egalitarian Curtailment is panel D in Figure \ref{fig:comparison-curtailments}.

\subsection{Summary of Unification}
These results show that the main fair curtailment schemes proposed in the literature \cite{gebbran2021fair, liu2020fairness} do not represent competing fairness axioms, but rather applications of the KS rule to different normative configurations of fallback and utopia references. Table~\ref{tab:KSsummary} summarizes the choices.

Depending on the problem at hand, the designer’s choice of these reference points ($u^0$ and $u^{\max}$) may be dictated by specific policy objectives, grid regulations, or the physical characteristics of the distribution network. For instance, choosing $d_i$ as a fallback acknowledges a prosumer's right to self-consumption, whereas a fallback based on $\bar{p}_i - \underline{c}$ prioritizes an egalitarian burden of curtailment volumes.

While we have unified these existing schemes under the generation metric $u_i(x) = x_i$, the welfarist framework allows for greater flexibility through the selection of alternative utility proxy metrics, thus expanding the designer's freedom to align curtailment with broader societal goals. Investigation of the optimal selection of these metrics is left for future work.

\begin{table}[t]
    \centering
    \caption{Existing fairness schemes as KS solutions with $u_i(x) = x_i$}
    \label{tab:KSsummary}
    \begin{tabular}{@{}lcc@{}}
        \toprule
        & Fallback $u^0_i$ & Utopia $u^{\max}_i$ \\
        \midrule 
        OPF Generation \cite{liu2020fairness} & 0 & $\bar p_i$\\
        Proportional redistr. \cite{gebbran2021fair}, OPF Export \cite{liu2020fairness} & $d_i$ & $\bar p_i$\\
        Uniform dynamic export \cite{gebbran2021fair} & $d_i$ & $d_i+K$ \\
        Egalitarian curtailment \cite{gebbran2021fair} & $\bar p_i -\underbar{c}$ & $\bar p_i$ \\
        \bottomrule
    \end{tabular}
\end{table}

\section{Simulations}\label{sec: simulations}

\begin{figure}[t]
    \centering
    \input{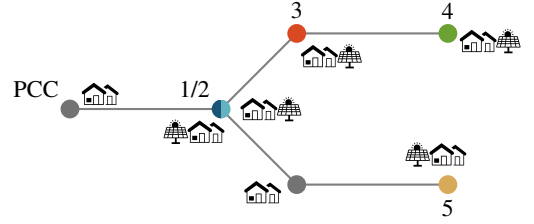}
    \caption{
    The 6-bus LV power grid testbed from the Swiss municipality of Walenstadt.
    The prosumers are numerically labeled.
    }
    \label{fig:wew-grid}
\end{figure}

The discussed curtailment schemes are evaluated on a 6-bus low-voltage power grid testbed representing a neighborhood in the Swiss municipality of Walenstadt (cf. Fig. \ref{fig:wew-grid}). The network comprises five prosumers connected at buses 1/2, 3, 4, and 5. The external medium-voltage grid is modeled as a slack bus with a fixed voltage magnitude of 1.046 p.u.

All simulations are performed using the balanced AC power flow solver of \texttt{pandapower} \cite{pandapower}. The reactive power is assumed to be zero at all buses, such that only active power injections from the prosumers are considered. Voltage magnitudes are constrained within 0.95 p.u. and 1.05 p.u. Artificial 24-hour demand and PV generation profiles are used, designed to reproduce a realistic duck-curve behavior. The profiles are normalized using the nominal PV capacities and demand values available for each bus, based on real system data.
The temporal resolution is 15 minutes, and curtailment decisions are applied independently at every time step.

We implement the four KS curtailment schemes from Section \ref{sec:unification} (reported in Table~\ref{tab:KSsummary}, as well as
\begin{itemize}
    \item the Nash product (see Section \ref{subsec:NashKS}) applied to export shares (corresponding to OF-Fair in \cite{Petrou2020}), and 
    \item the utilitarian SWF \eqref{eq:CFCform} with $\gamma = 0$ (corresponding to OPF Total in \cite{liu2020fairness}, OF-Total in \cite{Petrou2020}, Technical in \cite{Alam2024}, but also adopted in \cite{Dolan2014, Haque2016, Schermeyer2018}).
\end{itemize}
For the Nash and utilitarian formulations, the curtailment problem is solved using GUROBI, with a solution accuracy of $10^{-2}$ kW.

Figure \ref{fig:comparison-curtailments} shows a snapshot of the full time-series simulation at 12 p.m., corresponding to the period of peak PV generation.
Some observations are possible:
\begin{itemize}
    \item KS with $u_i^0 = 0,\ u_i^\text{max}=\bar p_i$ (panel A) -- prosumers are curtailed proportionally to their potential generation;
    \item KS with $u_i^0 = d_i,\ u_i^\text{max}=\bar p_i$ (panel B) -- prosumers' exports are curtailed in a proportionally fair way, self-consumption is considered a ``right'' that does not contribute to the decision;
    \item KS with $u_i^0 = d_i,\ u_i^\text{max}= d_i+K$ (panel C) -- prosumers enjoy the same ``right to export'' (in absolute terms) that does not depend on their self-consumption nor on their potential generation;
    \item KS with $u_i^0 = \bar p_i - \underbar{c},\ u_i^\text{max}= \bar p_i$ (panel D) -- prosumers are curtailed equally (in absolute terms), even if that affects their self-consumption;
    \item Utilitarian (panel E) -- prosumers are curtailed in the most efficient way (minimal total curtailment) even if that affects them very unequally, based on their electric distance from the substation;
    \item Nash (panel F) -- to maximize the product of exports, prosumers are curtailed rather unequally (again, based on their electric distance from the substation).
\end{itemize}

For further illustration, Fig. \ref{fig:24h} presents the complete 24-hour simulation results for the KS solution on export shares (Section \ref{subsec: opf_export_proportional}). The top panel shows the voltage magnitudes at the prosumer buses. The middle panel shows, for each prosumer, the available PV generation, the actual (curtailed) generation, and the demand. The bottom panel shows the evolution of the common export share $\lambda=(x_i - d_i)/(\bar{p}_i - d_i)$, which is granted uniformly to all prosumers.

Around the midday PV peak, the voltage at bus 4 reaches the upper feasibility boundary, indicating that the voltage constraints are marginally satisfied. This confirms that the computed values of $\lambda$ correspond to the KS solution, as it is the maximum feasible common export share that can be achieved without violating network constraints.

An interesting direction for future work consists in performing a similar social welfare optimization in an online setting, when the accumulated utility of the agents (that is, the curtailment that a prosumer is subject to over time) is considered \cite{Moring2024}.

\begin{figure}[t]
\centering
    \includegraphics[width=.8\linewidth,trim={0 1cm 0 0},clip]{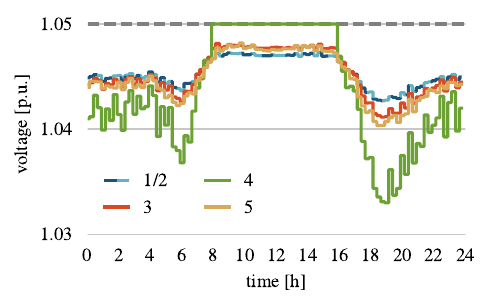}
    \\
    \vspace{.2cm}
    \includegraphics[width=.8\linewidth,trim={0 1cm 0 0},clip]{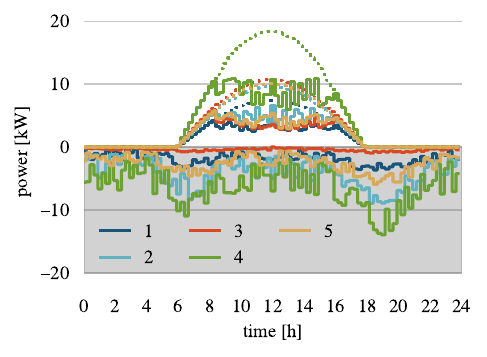}
    \\
    \vspace{.2cm}
    \includegraphics[width=.8\linewidth]{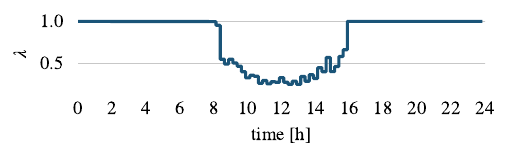}
\label{fig:24h}
\caption{24 hour profiles resulting from OPF Export alias KS on export shares. \textit{Top:} Voltages at the prosumer buses. \textit{Middle:} Available generation (dotted), actual generation after curtailment (solid positive), and demand (solid negative) for each prosumer. \textit{Bottom:} Common export share $\lambda$ granted to the prosumers.}
\end{figure}

\section{Conclusion}\label{sec: conclusion}
This work provides a principled and axiomatic foundation for fair generation curtailment, offering a rigorous alternative to ad-hoc metrics that can lead to system-wide inefficiencies. By modeling the operator's decision as an axiomatic-based Kalai-Smorodinsky social welfare function, we improve transparency of the design process: the choice of fallback and utopia reference points becomes an explicit and explainable design decision.

Our approach enjoys a high level of robustness because it does not assume interpersonal comparability of utilities, eliminating the need to estimate the economic value of power for heterogeneous consumers. Future research will investigate how additional information about the individual preferences should dictate the optimal selection of utility proxy metrics, and how specific social objectives can guide the definition of the reference benchmarks.

\section{AI Usage Disclosure}
AI tools were used during manuscript preparation to support the following: language editing (grammar and spelling) and refinement of draft text based on existing notes. All substantive content, results, and conclusions were developed and verified by the authors, who take full responsibility for the work.



\bibliographystyle{ieeetr}
\bibliography{references}
\balance

\end{document}